\documentclass[sigconf]{acmart}

\copyrightyear{2018} 
\acmYear{2018} 
\setcopyright{acmcopyright}
\acmConference[SIGIR '18]{The 41st International ACM SIGIR Conference on Research and Development in Information Retrieval}{July 8--12, 2018}{Ann Arbor, MI, USA}
\acmBooktitle{SIGIR '18: The 41st International ACM SIGIR Conference on Research and Development in Information Retrieval, July 8--12, 2018, Ann Arbor, MI, USA}
\acmPrice{15.00}
\acmDOI{10.1145/123_4}
\acmISBN{123-4567-24-567/18/07}
\usepackage[T1]{fontenc}
\usepackage{epstopdf}
\usepackage{multirow}
\usepackage{balance}
\usepackage{amsfonts}
\usepackage{amsmath}
\usepackage{amssymb}
\usepackage{graphicx}
\usepackage{booktabs} % For formal tables
\usepackage[noend,ruled,linesnumbered]{algorithm2e}
\usepackage{microtype}
\usepackage{url}
\usepackage{enumitem}
\usepackage{wrapfig,lipsum}
\usepackage{multirow}
\usepackage{makecell}
\usepackage{wrapfig}

\usepackage{subfig}
\usepackage{booktabs}
\usepackage{caption}

\usepackage{color,soul}

\newtheorem{thm:def}{Definition}
\newtheorem{thm:eg}{Example}
\newtheorem{thm:lem}{Lemma}
\newtheorem{thm:obs}{Observation}
\newtheorem{thm:req}{Requirement}

\newcommand{\nop}[1]{}

\newcommand{\mquote}[1]{{``\emph{#1}''}}
\newcommand{\SetRank}{\mbox{\sf SetRank}}
\newcommand{\SetRankFull}{\mbox{\sf SetRank}}
\newcommand{\SetRankNoType}{\mbox{\sf SetRank$^{-t}$}}
\newcommand{\SetRankNoTypeNoSet}{\mbox{\sf SetRank$^{-ts}$}}
\newcommand{\AutoSetRank}{\mbox{\sf AutoSetRank}}
\newcommand{\ie}{{\sl i.e.}}
\newcommand{\eg}{{\sl e.g.}}

\newcommand{\etal}{{\sl et al.}}

\DeclareMathAlphabet{\mathbbold}{U}{bbold}{m}{n}

\newcommand\rankeq{\mathrel{\overset{\makebox[0pt]{\mbox{\normalfont\tiny\sffamily rank}}}{=}}}
\newcommand\defeq{\mathrel{\overset{\makebox[0pt]{\mbox{\normalfont\tiny\sffamily def}}}{=}}}

\DeclareMathOperator*{\argmax}{arg\,max}

\begin{document}

\title{Entity Set Search of Scientific Literature: \\ An Unsupervised Ranking Approach}

\author{Jiaming Shen, Jinfeng Xiao, Xinwei He, Jingbo Shang, Saurabh Sinha, Jiawei Han}
\affiliation{
  \institution{Department of Computer Science, University of Illinois Urbana-Champaign, IL, USA}
}
\affiliation{
  \institution{ \{js2, jxiao13, xhe17, shang7, sinhas, hanj\}@illinois.edu }
}

\renewcommand{\shorttitle}{Entity Set Search of Scientific Literature: An Unsupervised Ranking Approach}

\begin{abstract}
    %!TEX root = main.tex
% UTF-8 encoding

Literature search is critical for any scientific research.
Different from Web or general domain search, a large portion of queries in scientific literature search are \emph{entity-set queries}, that is, \emph{multiple entities of possibly different types}.
Entity-set queries reflect user's need for finding documents that contain multiple entities and reveal inter-entity relationships and thus pose non-trivial challenges to existing search algorithms that model each entity separately.
However, entity-set queries are usually sparse (\ie, not so repetitive), which makes ineffective many supervised ranking models that rely heavily on associated click history.
To address these challenges, we introduce \SetRank, an unsupervised ranking framework that models inter-entity relationships and captures entity type information.
Furthermore, we develop a novel unsupervised model selection algorithm, based on the technique of weighted rank aggregation, to automatically choose the parameter settings in \SetRank~without resorting to a labeled validation set.
We evaluate our proposed unsupervised approach using datasets from TREC Genomics Tracks and \textsf{Semantic Scholar}'s query log.
The experiments demonstrate that \SetRank~significantly outperforms the baseline unsupervised models, especially on entity-set queries, and our model selection algorithm effectively chooses suitable parameter settings.

\end{abstract}

%\begin{CCSXML}
%<ccs2012>
%<concept>
%<concept_id>10002951.10003317.10003338</concept_id>
%<concept_desc>Information systems~Retrieval models and ranking</concept_desc>
%<concept_significance>300</concept_significance>
%</concept>
%</ccs2012>
%\end{CCSXML}
%\ccsdesc[300]{Information systems~Retrieval models and ranking}


\maketitle

%!TEX root = main.tex
% UTF-8 encoding
\section{Introduction}\label{sec:intro}

%% Why literature search
Literature search helps a researcher identify relevant papers and summarize essential claims about a topic, forming a critical step in any scientific research.
With the fast-growing volume of scientific publications, a good literature search engine is essential to researchers, especially in the domains like computer science and biomedical science where the literature collections are so massive, diverse, and rapidly evolving---few people can master the state-of-the-art comprehensively and in depth.

%% What is entity-set queries? Two concrete examples
A large set of literature search queries contain multiple entities which can be either concrete instances (\eg, \textit{GABP} (a gene)) or abstract concepts (\eg, \textit{clustering}).
We refer these queries as \emph{entity-set queries}.
For example, a computer scientist may want to find out how \textit{knowledge base} can be used for \textit{document retrieval} and thus issues a query ``\emph{knowledge base} for \emph{document retrieval}'', which is an entity-set query containing two entities.
Similarly, a biologist may want to survey how genes \textit{GABP}, \textit{TERT}, and \textit{CD11b} are associated with \textit{cancer} and submits a query ``\emph{GABP TERT CD11b cancer}'', another entity-set query with one disease and three gene entities.

%% What is special for entity-set queries? Two examples
Compared with typical short keyword queries, a distinctive characteristic of entity-set queries is that they reflect user's need for finding documents containing inter-entity relations.
For example, among 50 queries collected from biologists in 2005 as part of TREC Genomics Track \cite{Hersh2005TREC2G}, 40 of them are explicitly formulated as finding relations among at least two entities.
In most cases, a user who submits an entity-set query will expect to get a ranked list of documents that are most relevant to the whole \emph{entity set}.
Therefore, as in the previous examples, returning a paper about only \textit{knowledge bases} or only one gene \textit{GABP} is unsatisfactory.

%% Table: ESQ performance on S2 dataset
\small
\begin{table}[!t]
\caption{Ranking performance on 100 benchmark queries of the \textsf{S2} production system.  Entity-set queries (ESQs), marked \textsf{\textbf{bold}}, perform much weaker than non-ESQs do.}\label{tbl:s2-query-performance}
\vspace{-0.3cm}
            \begin{tabular}{|c|c|c|c|c|}
                \hline
                Metrics & ESQs   & non-ESQs & Overall  \\
                \hline
                \hline
                NDCG@5 & \textbf{0.3622}  &  0.6291 & 0.5223  \\
                \hline
                NDCG@10 & \textbf{0.3653} & 0.6286 & 0.5233  \\
                \hline
                NDCG@15 & \textbf{0.3840} & 0.6221 & 0.5269  \\
                \hline
                NDCG@20 & \textbf{0.4011}  & 0.6247 & 0.5353   \\
                \hline
            \end{tabular}
            \vspace{-0.2cm}
\end{table}
\normalsize

%% Why previous methods do not work well for entity-set queries. What are the challenges?
Entity-set queries pose non-trivial challenges to existing search platforms.
For example, among the 100 queries\footnote{\url{http://data.allenai.org/esr/Queries/}} released by \textsf{Semantic Scholar (S2)}, 40 of them are entity-set queries and \textsf{S2}'s production ranking system performs poorly on these entity-set queries, as shown in Table \ref{tbl:s2-query-performance}.
The difficulties of handling entity-set queries mainly come from two aspects.
First, entity relations within entity sets have not been modeled effectively.
The association or co-occurrence of multiple entities has not gained adequate attention from existing ranking models.
As a result, those models will rank papers where a single distinct entity appears multiple times higher than those containing many distinct entities.
Second, entity-set queries are particularly challenging for supervised ranking models.
As manual labeling of document relevance in academic search requires domain expertise, it is too expensive to train a ranking model based purely on manually labeling.
Most systems will first apply an off-the-shelf unsupervised ranking model during their \emph{cold-start} process and then collect user interaction data (\eg, click information).
Unfortunately, entity-set queries are usually sparse (\ie, not so repetitive), and have less associated click information.
Furthermore, many off-the-shelf unsupervised models cannot return reasonably good candidate documents for entity-set queries within the top-20 positions.
Many highly relevant documents will not be presented to users, which further compromises the usefulness of clicking information.

%% Following are our methodology part
This paper tackles the new challenge---\textit{improving the search quality of scientific literature on entity-set queries} and proposes an unsupervised ranking approach.
We introduce \SetRank, an unsupervised ranking framework that explicitly models inter-entity relations and captures entity type information.
\SetRank~first links entity mentions in query and documents to an external knowledge-base.
Then, each document is represented with both bag-of-words and bag-of-entities representations \cite{Xiong2016BagofEntitiesRF, Xiong2017WordEntityDR} and fits two language models respectively.
On the query side, a novel heterogeneous graph representation is proposed to model complex entity information (\eg, entity type) and entity relations within the set.
This heterogeneous query graph represents all the information need in that query.
Finally, the query-document matching is defined as a \emph{graph covering process} and each document is ranked based on the information need it covers in the query graph.

Although being an unsupervised ranking framework, \SetRank~still has some parameters that need to be appropriately learned using a labeled validation set.
To further automate the process of ranking model development, we develop a novel unsupervised model selection algorithm based on the technique of weighted rank aggregation.
Given a set of queries with no labeled documents, and a set of candidate parameter settings, this algorithm automatically learns the most suitable parameter settings for that set of queries.

The significance of our proposed unsupervised ranking approach is two-fold.
First, \SetRank~itself, as an unsupervised ranking model, boosts the literature search performance on entity-set queries.
Second, \SetRank~can be adopted during the \emph{cold-start} process of a search system, which enables the collection of high-quality click data for training subsequent supervised ranking model.
Our experiments on \textsf{S2}'s benchmark datasets and TREC 2004 \& 2005 Genomics Tracks \cite{Hersh2004TREC2G, Hersh2005TREC2G} demonstrate the usefulness of our unsupervised model selection algorithm and the effectiveness of \SetRank~for searching scientific literature, especially on entity-set queries.

%% Contributions
In summary, this work makes the following contributions:
\vspace*{-1.0ex}
\begin{enumerate}[leftmargin=*]
\item A new research problem, effective entity-set search of scientific literature, is studied.
\item \SetRank, an unsupervised ranking framework, is proposed, which models inter-entity relations and captures entity type information.
\item A novel unsupervised model selection algorithm is developed, which automatically selects \SetRank's parameter settings without resorting to a labeled validation set.
\item Extensive experiments are conducted in two scientific domains, demonstrating the effectiveness of \SetRank~and our unsupervised model selection algorithm.
\end{enumerate}
\vspace*{-1.0ex}

%% Paper Organization
The remaining of the paper is organized as follows.
Section \ref{sec:related} discusses related work.
Section \ref{sec:ranking_framework} presents our ranking framework \SetRank.
Section \ref{sec:rank_aggregation} presents the unsupervised model selection algorithm.
Section \ref{sec:exp} reports and analyzes the experimental results on two benchmark datasets and shows a case study of \SetRank~for biomedical literature search.
Finally, Section \ref{sec:con} concludes this work with discussions on some future directions.

%!TEX root = main.tex
% UTF-8 encoding

\section{Related Work}\label{sec:related}

We examine related work in three aspects: academic search, entity-aware ranking model, and automatic ranking model selection.

%% Academic Search
\vspace{-0.2cm}
\subsection{Academic Search}
The practical importance of finding highly relevant papers in scientific literature has motivated the development of many academic search systems.
Google Scholar is arguably the most widely used system due to its large coverage.
However, the ranking result of Google Scholar is still far from satisfactory because of its bias toward highly cited papers \cite{Beel2009GoogleSR}.
As a result, researchers may choose other academic search platforms, such as CiteSeerX \cite{Wu2014CiteSeerXAI}, AMiner \cite{Tang2008ArnetMinerEA}, PubMed \cite{Lu2011PubMedAB}, Microsoft Academic Search \cite{Sinha2015AnOO} and Semantic Scholar \cite{Xiong2017ExplicitSR}.
Research efforts of many such systems focus on the analytical tasks of scholar data such as author name disambiguation \cite{Tang2008ArnetMinerEA}, paper importance modeling \cite{Shen2016ModelingTA}, and entity-based distinctive summarization \cite{Ren2017LifeiNetAS}.
However, this work focuses on ad-hoc document retrieval and ranking in academic search.
The most relevant work to ours is \cite{Xiong2017ExplicitSR} in which entity embeddings are used to obtain ``soft match'' feature of each $\langle$query, document$\rangle$ pair.
However, \cite{Xiong2017ExplicitSR} requires training data to combine word-based and entity-based relevance scores and to select parameter settings, which is rather different from our unsupervised approach. 

%% Entity-aware Ranking Model
\vspace{-0.2cm}
\subsection{Entity-aware Ranking Model}
Entities, such as people, locations, or abstract concepts, are natural units for organizing and retrieving information \cite{Garigliotti2017OnTE}.
Previous studies found that over 70\% of Bing's query and more than 50\% of traffic in \textsf{Semantic Scholar} are related to entities \cite{Guo2009NamedER, Xiong2017ExplicitSR}.
The recent availability of large-scale knowledge repositories and accurate entity linking tools have further motivated a growing body of work on entity-aware ranking models.
These models can be roughly categorized into three classes: expansion-based, projection-based, and representation-based.

The expansion-based methods use entity descriptions from knowledge repositories to enhance query representation.
Xu \etal~\cite{Xu2009QueryDP} use entity descriptions in Wikipedia as pseudo relevance feedback corpus to obtain cleaner expansion terms;
Xiong and Callen \cite{Xiong2015QueryEW} utilize the description of Freebase entities related to the query for query expansion;
Dalton \etal~\cite{Dalton2014EntityQF} expand a query using the text fields of the attributes of the query-related entities and generate richer learning-to-rank features based on the expanded texts.

The projection-based methods try to project both query and document onto an entity space for comparison.
Liu and Fang \cite{Liu2015LatentES} use entities from a query and its related documents to construct a latent entity space and then connect the query and documents based on the descriptions of the latent entities.
Xiong and Callen \cite{Xiong2015EsdRankCQ} use the textual features among query, entities, and documents to model the query-entity and entity-document connections.
These additional connections between query and document are then utilized in a learning-to-rank model.
A fundamental difference of our work from the above methods is that we do not represent query and document using external terms/entities that they do not contain.
This is to avoid adding noisy expansion of terms/entities that may not reflect the information need in the original user query.

The representation-based methods, as a recent trend for utilizing entity information, aim to build entity-enhanced text representation and combine it with traditional word-based representation \cite{Xiong2017WordEntityDR}.
Xiong \etal~\cite{Xiong2016BagofEntitiesRF} propose a bag-of-entities representation and demonstrated its effectiveness for vector space model.
Raviv \etal~\cite{Raviv2016DocumentRU} leverage the surface names of entities to build an entity-based language model.
Many supervised ranking models are proposed to apply learning-to-rank methods for combining entity-based signals with word-based signals.
For example, ESR \cite{Xiong2017ExplicitSR} uses entity embeddings to compute entity-based query-document matching score and then combines it with word-based score using RankSVM.
Following the same spirit, Xiong \etal~\cite{Xiong2017WordEntityDR} propose a word-entity duet framework that simultaneously models the entity annotation uncertainty and trains the ranking model.
Comparing with the above methods, we also use the bag-of-entity representation but combine it with word-based representation in an unsupervised way.
Also, to the best of our knowledge, we are the first to capture entity relation and type information in an unsupervised entity-aware ranking model.

%% Automatic Ranking Model Selection
\vspace{-0.2cm}
\subsection{Automatic Ranking Model Selection}
Most ranking models need to manually set many parameter values.
To automate the process of selecting parameter settings, some AutoML methods \cite{Feurer2015EfficientAR, Brazdil2017MetalearningAA} are proposed.
Nevertheless, these methods still require a validation set which contains queries with labeled documents.
In this paper, we develop an unsupervised model selection algorithm, based on rank aggregation, to automatically choose parameter settings without resorting to a labeled validation set.
Rank aggregation aims to combine multiple existing rankings into a joint ranking.
Fox and Shaw \cite{Fox1993CombinationOM} propose some deterministic functions to combine rankings heuristically.
Klementiev \etal~\cite{Klementiev2007AnUL, Klementiev2008AFF} propose an unsupervised learning algorithm for rank aggregation based on a linear combination of ranking functions.
Another related line of work is to model rankings using a statistical model (\eg, Plackett-Luce model) and aggregate them based on statistical inference \cite{Guiver2009BayesianIF, Maystre2015FastAA, Zhao2016LearningMO}.
Lately, Bhowmik and Ghosh \cite{Bhowmik2017LETORMF} propose to use object attributes to augment some standard rank aggregation framework.
Compared with these methods, our proposed algorithm goes beyond just combining multiple rankings and uses aggregated ranking to guide the selection of parameter settings.

%!TEX root = main.tex
% UTF-8 encoding
\section{Ranking Framework}\label{sec:ranking_framework}

% Figure: Document representation
\begin{figure}[t]
  \centering
  \includegraphics[width=0.48\textwidth]{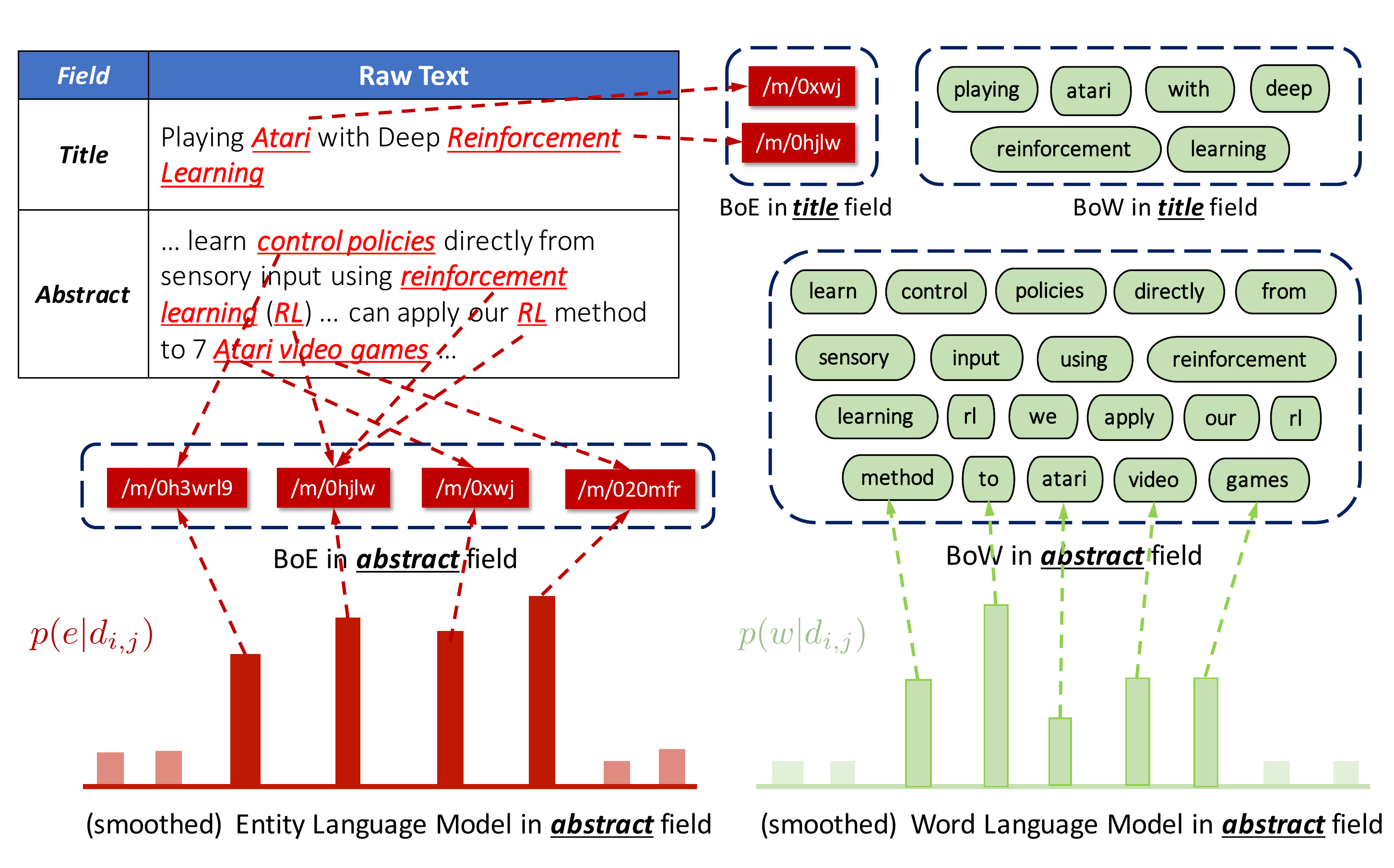}
  \caption{\small An illustrative example showing one document comprised of two fields (\ie, title, abstract) with their corresponding bag-of-words and bag-of-entities representations.}
  \label{fig:DocumentRepresentation}
  \vspace{-0.1cm}
\end{figure}

This section presents our unsupervised ranking framework for leveraging entity (set) information in search. 
Our framework provides a principled way to rank a set of documents $D$ for a query $q$. 
In this framework, we represent each document using standard bag-of-words and bag-of-entities representations~\cite{Xiong2016BagofEntitiesRF,Xiong2017WordEntityDR} (Section~\ref{subsec:document_representation}) and represent the query using a novel heterogeneous graph (Section~\ref{subsec:query_representation}) which naturally model the entity set information.
Finally, we model the query-document matching as a ``graph covering" process, as described in Section~\ref{subsec:ranking_model}.

%% Document Representation
\subsection{Document Representation}\label{subsec:document_representation}

We represent each document using both word and entity information.
For words, we use standard bag-of-words representation and treat each unigram as a word.
For entities, we adopt an entity linking tool (details described in Section 5.2) that utilizes a knowledge base/graph (\eg, Wikidata or Freebase) where entities have unique IDs.
Given an input text, this tool will find the entity mentions (\ie, entity surface names) in the text and link each of them to a disambiguated entity in the knowledge base/graph. 
For example, given the input document title \mquote{Training linear \underline{SVMs} in linear time}, this tool will link the entity mention \mquote{SVMs}' to the entity \mquote{Support Vector Machine} with Freebase id \textsf{`/m/0hc2f'}.
Previous studies \cite{Xiong2016BagofEntitiesRF,Raviv2016DocumentRU} show that when the entity linking error is within a reasonable range, the returned entity annotations, though noisy, can still improve the overall search performance, partially due to the following:
\begin{enumerate}[leftmargin=*]
\item \textbf{Polysemy resolution.} Different entities with the same surface name will be resolved by the entity linker. For example, the fruit \mquote{Apple} (with id \textsf{`/m/014j1m'}) will be disambiguated with the company \mquote{Apple} (with id \textsf{`/m/0k8z'}).
\item \textbf{Synonymy resolution.} Different entity surface names corresponding to the same entity will be identified and merged. For example, the entity \mquote{United States of America} (with id \textsf{`/m/09c7w0'}) can have different surface names including \mquote{USA}, \mquote{United States}, and \mquote{U.S.} \cite{Raviv2016DocumentRU}. The entity linker can map all these surface names to the same entity.
\end{enumerate}

After linking all the entity mentions in a document to entities in the knowledge base, we can obtain the bag-of-entities representation of this document.
Then, we fit two language models (LMs) for this document: one being word-based (\ie, traditional unigram LM) and the other being entity-based.
Notice that in the literature search scenario, documents (\ie, papers) usually contain multiple fields, such as title, abstract, and full text.
We model each document field using a separate bag-of-words representation and a separate bag-of-entities representation, as shown in Figure \ref{fig:DocumentRepresentation}.

To exploit such intra-document structures, we generally assume a document $d_{i}$ has $k$ fields $d_{i} = \{d_{i,1}, \dots, d_{i,k} \}$ and thus the document collection can be separated into $k$ parts: $\{ D_{1}, \dots, D_{k} \}$. 
Following \cite{Ogilvie2003CombiningDR}, we assign each field a weight $\delta_{j}$ and formulate the generation process of a token $t$ given the document $d_{i}$ as follows:
\small
\begin{equation}
p(t|d_{i}) = \sum_{j=1}^{k} p(t|d_{i,j})p(d_{i,j} | d_{i}), \quad p(d_{i,j} | d_{i}) = \frac{\delta_{j}}{\sum_{j'=1}^{k} \delta_{j'}}.
\end{equation}
\normalsize
Notice the a token $t$ can be either a unigram $w$ or an entity $e$, and the field weight $\delta_{j}$ can be either manually set based on prior knowledge or automatically learned using the mechanism described in Section~\ref{sec:rank_aggregation}.
The token generation probability under each document field $p(t|d_{i,j})$ can be obtained from the maximum likelihood estimate with Dirichlet prior smoothing~\cite{Zhai2001ASO} as follows:
\small
\begin{equation}\label{eq:lm}
p(t|d_{i,j}) = \frac{n_{t,d_{i,j}} + \mu_{j} \frac{n_{t, D_{j}}}{L_{D_{j}}}}{L_{d_{i,j}} + \mu_{j}},
\end{equation}
\normalsize
where $n_{t, d_{i,j}}$ and $L_{d_{i,j}}$ represent the number of token $t$ in $d_{i,j}$ and the length of $d_{i,j}$. 
Similarly, we can define $n_{t, D_{j}}$ and $L_{D_{j}}$.
Finally, $\mu_{j}$ is a scale parameter of the Dirichlet distribution for field $j$.
A concrete example is shown in Figure \ref{fig:DocumentRepresentation}.

%% Query Representation
\subsection{Query Representation}\label{subsec:query_representation}

\begin{figure}[!t]
  \centering
  \includegraphics[width=0.48\textwidth]{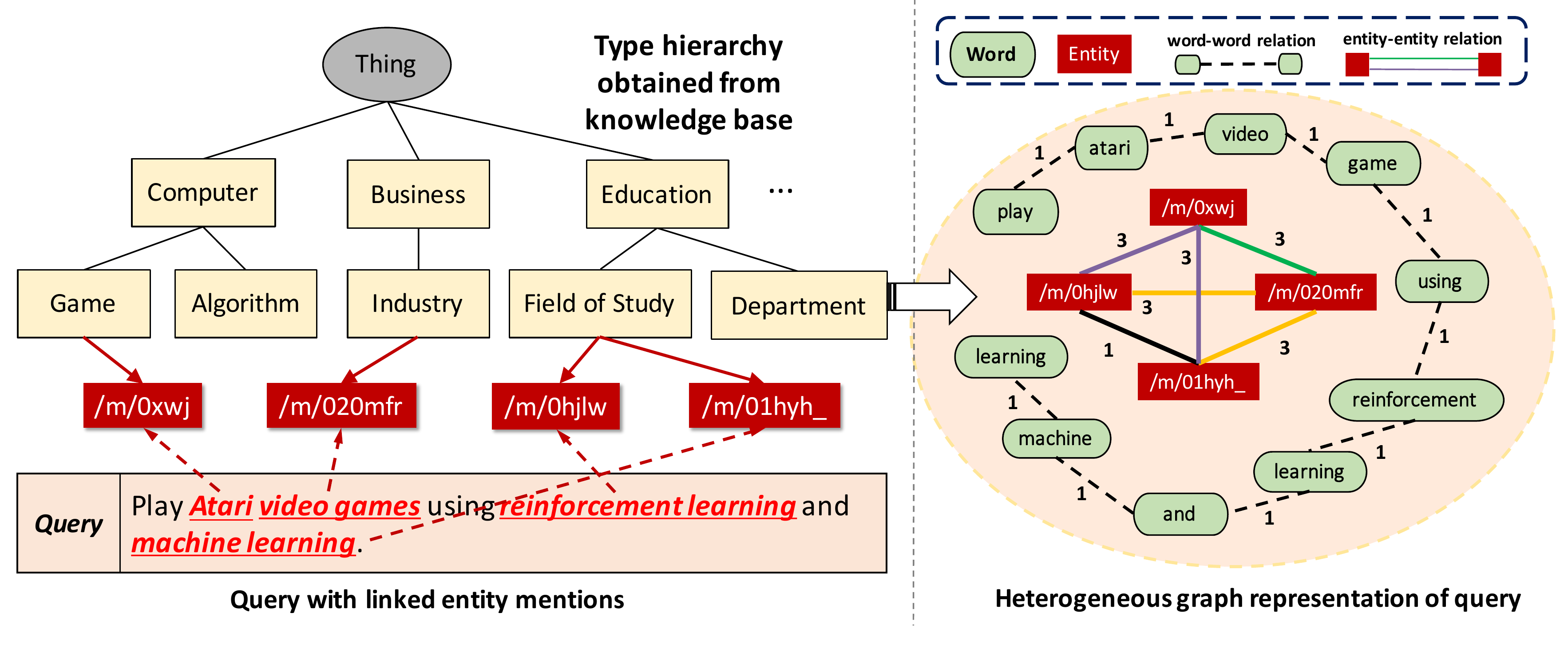}
  \caption{\small An illustrative example showing the heterogeneous graph representation of one query. Word-word relations are marked by dash lines and entity-entity relations are marked by solid lines. Different solid line colors represent different relation strengths based on two entities' types.}\label{fig:QueryRepresentation}
    \vspace{-0.1cm}
\end{figure}

Given an input query $q$, we first apply the same entity linker used for document representation to extract all the entity information in the query.
Then, we design a novel heterogeneous graph to represent this query $q$, denoted as $G_{q}$.
Such a graph representation captures both word and entity information in the query and models the entity relations.
A concrete example is shown in Figure \ref{fig:QueryRepresentation}.

\noindent
\textbf{Node representation.} % Node representation
In this heterogeneous query graph, each node represents a query token. As a token can be either a word or an entity, there are two different types of nodes in this graph.

\noindent
\textbf{Edge representation.} % Edge representation
We use an edge to represent a latent relation between two query tokens. In this work, we consider two types of latent relations: word-word relation and entity-entity relation.
For word-word relation, we add an edge for each pair of adjacent word tokens with equal weight 1.
For instance, given an query \mquote{Atari video games}, we will add two edges, one between word pairs \emph{$\langle$Atari, video$\rangle$} and the other between \emph{$\langle$video, game$\rangle$}.
On the entity side, we aim to emphasize all the possible entity-entity relations, and thus add an edge between each pair of entity tokens.

\noindent
\textbf{Modeling entity type.} % Modeling entity type
The type information of each query entity can further reveal the user's information need. 
Therefore, we assign the weight of each entity-entity relation based on these two entities' type information.
Intuitively, if the types of two entities are distant from each other in a type hierarchy, then the relation between these two entities should have a larger weight.
A similar idea is exploited in \cite{Garigliotti2017OnTE} and found useful for type-aware entity retrieval.

Mathematically, we use $\phi_e$ to denote the type of entity $e$; use $LCA_{u, v}$ to denote the Lowest Common Ancestor (LCA) of two nodes $u$ and $v$ in a given tree (\ie, type hierarchy), and use $l(u, v)$ to denote the length of a path between node $u$ and node $v$.
In Figure \ref{fig:QueryRepresentation}, for example, the entity tokens \textsf{`/m/0hjlw'} and \textsf{`/m/0xwj'}, corresponding to \mquote{reinforcement learning} and \mquote{Atari}, have types \textsf{`education.field\_of\_study'} and \textsf{`computer.game'}, respectively.
The Lowest Common Ancestor of these two types in the type hierarchy is \textsf{`Thing'}.
Finally, we define the relation strength between entity $e_{1}$ and entity $e_{2}$ as follows:
\small
\begin{eqnarray}\label{eq:edge-weight}
LCA_{e_1, e_2} & = & LCA(\phi_{e_1}, \phi_{e_2}), \\
\lambda_{e_1, e_2} & = & 1 + \max\left\{ l(\phi_{e_1}, LCA_{e_1, e_2} ) , l(\phi_{e_2}, LCA_{e_1, e_2} )\right\}.
\end{eqnarray}
\normalsize
Our proposed heterogeneous query graph representation is general and can be extended.
For example, we can apply dependency parsing for verbose queries, and only add an edge between two word tokens that have direct dependency relation.
Also, if the importance of each entity-entity relation is given, we can then set the edge weights correspondingly.
We leave these extensions for future works.

%% Figure: Query Representation

%% Document Ranking using Query Graph
\vspace{-0.1ex}
\subsection{Document Ranking using Query Graph}\label{subsec:ranking_model}
Our proposed heterogeneous query graph $G_{q}$ represents all information need in the user-issued query.
Such need can be either to find document discussing one particular entity or to identify papers studying an important inter-entity relation.
Intuitively, a document that can satisfy more information need should be ranked at a higher position.
To quantify such information need that is explained by a document, we define the following graph covering process.

\noindent \textbf{Query graph covering.}  % Query graph covering.
If a query token $t \in q$ exists in a document $d_i$, we say $d_i$ \emph{covers} the node in $G_q$ that corresponds to this token.
Similarly, if a pair of query tokens $t_1$ and $t_2$ exists in $d_i$, we say $d_i$ \emph{covers} the edge in $G_q$ that corresponds to the relation of this token pair $\langle t_1, t_2 \rangle$.
The subgraph of $G_{q}$ that is covered by the document $d_{i}$, denoted as $G_{q|d_i}$, represents the information need in the query $q$ that is explained by the document $d_i$.

Furthermore, we follow the same spirit of \cite{metzler2005markov} and view the subgraph $G_{q|d_i}$ as a Markov Network, based on which we define the joint probability of the document $d_{i}$ and the query $q$ as follows:
\small
\begin{equation}
\label{eq:MRFranking}
    P(d_{i}, q) ~\defeq~ \frac{1}{Z} \prod_{c \in G_{q|d_{i}}} \psi(c) ~\rankeq~  \sum_{c \in G_{q|d_{i}}} \log \psi(c)  ~\rankeq~  \sum_{c \in G_{q|d_i}} f(c),
\end{equation}
\normalsize
where $Z$ is a normalization factor, $c$ indexes the cliques in graph, and $\psi(c)$ is the non-negative potential defined on $c$. The last equation holds as we let $\psi(c) = \exp[f(c)]$.
Notice that if $G_{q|d_1}$ is a subgraph of $G_{q|d_2}$ which means document $d_{1}$ covers \emph{less} information than document $d_{2}$ does, we should have $P(d_{1}, q) < P(d_{2}, q)$.
Therefore, we should design $f(\cdot)$ to satisfy the constraint $f(c) > 0, \forall c$.

In this work, we focus on modeling each single entity and pairwise relations between two entities.
Therefore, each clique $c$ can be either a node or an edge in the graph.
Modeling higher-order relations among more than two entities (\ie, cliques with size larger than 2) is left for future work.
We define the potential functions for a single node and an edge as follows:

\vspace{0.1cm}
\noindent \textbf{Node potential.} % Node potential
Node potential quantifies the information need contained in a single node $t$, which can be either a word token $w$ or an entity token $e$.
To balance the relative weight of a word token and an entity token, we introduce a parameter $\lambda_{E} \in [0,1]$, and define the node potential function $f(\cdot)$ as follows:
\small
\begin{equation}\label{eq:node_lambda}
f(t)   =  \left\{ \begin{array}{ll}
				\lambda_{E} \cdot a(P(t|d_{i}))  & \textrm{if token $t$ is an entity token} \\
				(1-\lambda_{E}) \cdot a(P(t|d_{i}))  & \textrm{if token $t$ is a word token} \\
			\end{array}
			\right.
\end{equation}
\normalsize
where $a(\cdot)$ is an activation function that transforms a raw probability to a node potential. Here, we set $a(x) = \sqrt{x}$ in order to amplify $P(t|d_{i})$ which has a relatively small value.

\vspace{0.1cm}
\noindent \textbf{Edge potential.}  % Edge potential
Edge potential quantifies the information need contained in an edge $\langle t_1, t_2 \rangle$ that can be either a word-word (W-W) relation or and an entity-entity (E-E) relation.
In our query graph representation, all word-word relations have an equal weight of 1, and the weight of each entity-entity relation (\ie, $\lambda_{e_1, e_2}$) is defined by Equation (\ref{eq:edge-weight}).
Finally, we calculate the edge potential as follows:
\small
\begin{equation}\label{eq:edge-potential}
f(\langle t_1,t_2\rangle) = \lambda_{\langle t_1, t_2\rangle} \cdot a(P(t_1, t_2 |d_{i})),
\end{equation}
\begin{equation}\label{eq:edge_lambda}
\lambda_{\langle t_1, t_2\rangle}    =  \left\{ \begin{array}{ll}
				\lambda_{E} \cdot \lambda_{e_1, e_2} & \textrm{if $\langle t_1, t_2 \rangle$ is an E-E relation} \\
				(1-\lambda_{E})  & \textrm{if $\langle t_1, t_2 \rangle$ is a W-W relation} \\
			\end{array}
			\right.
\end{equation}
\normalsize
where $\lambda_{\langle t_1, t_2 \rangle}$ measures the edge importance, and $a(\cdot)$ is the same activation function as defined above.
To simplify the calculation of $P(t_1, t_2 |d_{i})$, we make an assumption that two tokens $t_1$ and $t_2$ are conditionally independent given a document $d_{i}$.
Then, we replace $P(t_1, t_2 | d_{i})$ with $P(t_1 | d_{i})P(t_2 | d_{i})$ and substitute it in Equation (\ref{eq:edge-potential}).

%%% Summary
\vspace{0.1cm}
\noindent \textbf{Putting all together.}
After defining the node and edge potentials, we can calculate the joint probability of each document $d_{i}$ and query $q$ using Equation (\ref{eq:MRFranking}) as follows:
\small
\begin{equation}
\begin{aligned}
P(d_i, q) &= (1-\lambda_{E}) \sum_{w \in G_{q|d_{i}}} \left(  1 + \sum_{\langle w, w' \rangle \in G_{q|d_{i}}}  a(P(w'|d_{i}))    \right) a(P(w|d_{i}))  \\
	 &+ \lambda_{E} \sum_{e \in G_{q|d_{i}}}  \left(  1 + \sum_{\langle e, e' \rangle \in G_{q|d_{i}}}  \lambda_{e, e'} \cdot a(P(e'|d_{i}))    \right) a(P(e|d_{i})).
\end{aligned}
\end{equation}
\normalsize
As shown in the above equation, \SetRank~will explicitly reward paper capturing inter-entity relations and covering more unique entities.
Also, it uses $\lambda_{E}$ to balance the word-based relevance with entity-based relevance, and models entity type information in $\lambda_{e,e'}$.

%!TEX root = main.tex
% UTF-8 encoding
\section{Unsupervised Model Selection}\label{sec:rank_aggregation}

Although being an unsupervised ranking framework, \SetRank~still has some parameters that need to be appropriately set by ranking model designers, including the weight of title/abstract field and the relative importance of entity token $\lambda_{E}$. 
Previous study \cite{Zhai2001ASO} shows that these model parameters have significant influences on the ranking performance and thus we need to choose them carefully.
Typically, these parameters are chosen to optimize the performance over a validation set that is manually constructed and contains the relevance label of each query-document pair. 
Though being useful, the validation set is not always available, especially for those applications (\eg, literature search) where labeling document requires domain expertise.

To address the above problem, we propose an unsupervised model selection algorithm which automatically chooses the parameter settings without resorting to a manually labeled validation set. 
The key philosophy is that although people who design the ranking model (\ie, ranking model designers) do not know the exact ``optimal'' parameter settings, they do have prior knowledge about the reasonable range for each of them. 
For example, the title field weight should be set larger than the abstract field weight, and the entity token weight $\lambda_{E}$ should be set small if the returned entity linking results are noisy.
Our model selection algorithm leverages such prior knowledge by letting the ranking model designer input the search range of each parameter's value. 
It will then return the best value for each parameter within its corresponding search range.
We first describe our notations and formulate our problem in Section \ref{subsec:rank_aggregation_notation_problem}. 
Then, we present our model selection algorithm in Section \ref{subsec:rank_aggregation_model_selection}.

\subsection{Notations and Problem Formulation}\label{subsec:rank_aggregation_notation_problem}

\noindent \textbf{Notations.}
We use $S_{K}$ to denote the collection of rankings over a set of $K$ documents: $D = \{d_1, \dots, d_{k}, \dots, d_{K} \}$, $k \in [K] = \{1, \dots, K\}$.
We denote by $\pi: [K] \rightarrow [K]$ a \emph{complete ranking}, where $\pi(k)$ denotes the position of document $d_{k}$ in the ranking, and $\pi^{-1}(j)$ is the index of the document on position $j$. 
For example, given the ranking: $d_3 \succ d_1 \succ d_2 \succ d_4$, we will have $\pi = [2, 3, 1, 4]$ and $\pi^{-1} = (3, 1, 2, 4)$. 
Furthermore, we use the symbol $\tau$ (instead of $\pi$) to denote an \emph{incomplete ranking} which includes only some of the documents in $D$. 
If document $d_k$ does not occur in the ranking, we set $\tau(k) = 0$, otherwise, $\tau(k)$ is the rank of document $d_k$. 
In the corresponding $\tau^{-1}$, those missing documents simply do not occur. 
For example, given the ranking: $d_4 \succ d_2 \succ d_1$, we have $\tau = [3, 2, 0, 1]$ and $\tau^{-1} = (4, 2, 1)$. 
Finally, we let $I(\tau) = \{k | \tau(k) > 0, k \in [K] \} $ to represent the index of documents that appear in the ranking list $\tau$. 

\vspace{0.1cm}
\noindent \textbf{Problem Formulation.}
Given a parameterized ranking model $M_{\theta}$ where $\theta$ denotes the set of all parameters (\eg, $\{k, b\}$ in BM25, $\{\mu\}$ in query likelihood model with dirichlet prior smoothing), we want to find the best parameter settings $\theta^{*}$ such that the ranking model $M_{\theta^{*}}$ achieves the best ranking performance over the space $\mathbb{Q}$ of all queries. 
In practice, however, the space consisting of all possible values of $\theta$ can be infinite and we cannot access all queries in $\mathbb{Q}$. 
Therefore, we assume ranking model designers will input $p$ possible sets of parameter values: $\Theta = \{\theta_{1}, \dots, \theta_{p} \}$ and a finite subset of queries $Q \subset \mathbb{Q}$. 
Finally, we formulate our problem of \emph{unsupervised model selection} as follows:
%% Problem Formulation
\vspace{-0.5em}
\begin{thm:def}
(PROBLEM FORMULATION). Given a parameterized ranking model $M_{\theta}$, $p$ candidate parameter settings $\Theta$, and an unlabeled query subset $Q$, we aim to find $\theta^{*} \in \Theta$ such that $M_{\theta^{*}}$ achieves the best ranking performance over $Q$. 
\end{thm:def}
\vspace{-1.0em}

%We use the following concrete example to further elaborate our problem:
%\vspace{-0.5em}
%\begin{thm:eg}
%Suppose a biomedical researcher wants to build a literature search engine of publications about the cardiovascular disease. 
%He collects a subset of search queries $Q$ about the cardiovascular disease, and ``cold-starts" the system building using \SetRank.
%\end{thm:eg}
\subsection{Model Selection Algorithm}\label{subsec:rank_aggregation_model_selection}

%% Algorithm: Unsupervised Model Selection
\small
\begin{algorithm}[!t]
  \caption{Unsupervised Model Selection.}
  \label{alg:unsupervised_model_selection}
  \KwIn{
    A parameterized ranking model $M_{\theta}$, $p$ candidate parameter settings $\Theta = \{\theta_{1}, \cdots, \theta_{p} \}$, and an unlabeled query subset $Q$.
  }
  \KwOut{The best ranking model $M_{\theta^{*}}$ with $\theta^{*} \in \Theta$.}
  set $score(M_{\theta_{1}}) = score(M_{\theta_{2}}) = \cdots = score(M_{\theta_{p}}) = 0$\;
  \For{query $q \in Q$} {
  	set $\alpha_{1} = \alpha_{2} = \cdots \alpha_{p} = \frac{1}{p}$\;
	set $\pi^{prev} = None$\;
	\While{True} {
		//  Weighted Rank Aggregation \; 
  		\For{document index $j$ from 1 to $|D|$  }  { 
  			$score(d_j) = 0$\;
  			\For{ rank list index $i$ from 1 to $p$} { 
				\If{$j \in I(\tau_{i})$ (\ie, $d_{j}$ appears in $\tau_{i}$)} { 
					$score(d_j) = score(d_j) + \alpha_{i} (|\tau_{i}|+1-\tau_{i}(d_j))$\; 
				}
			}
  		}
		 $\pi = \text{argsort} ( score({d_1}), \cdots, score({d_{|D|}}) )$\;
		 //  Confidence Score Adjustment \; 
		 \For{rank list index $i$ from 1 to $p$  }  { 
  			$\alpha_{i} = \frac{\exp(-dist(\tau_{i} || \pi))}{\sum_{i'} \exp(-dist(\tau_{i'} || \pi)) }$\;
  		}
		//  Convergence Check \; 
		\If{$\pi == \pi^{prev}$} {
		 	Break;
		 }
		 \Else {
		 	$\pi^{prev} \gets \pi$\;
		 }
	}
	 \For{rank list index $i$ from 1 to $p$  }  { 
	 	$score(M_{\theta_{i}}) = score(M_{\theta_{i}}) + \alpha_{i}$\;
  	}
  }
  $M_{\theta^{*}} = \argmax_{\theta \in \Theta} score(M_{\theta})$\;
  Return $M_{\theta^{*}}$\;
\end{algorithm}
%\vspace{-0.5cm}
\normalsize

%% Figure: Illustrative example for weighted rank aggregation and confidence score adjustment. 
\begin{figure}[!t]
  \centering
  \includegraphics[width=0.46\textwidth]{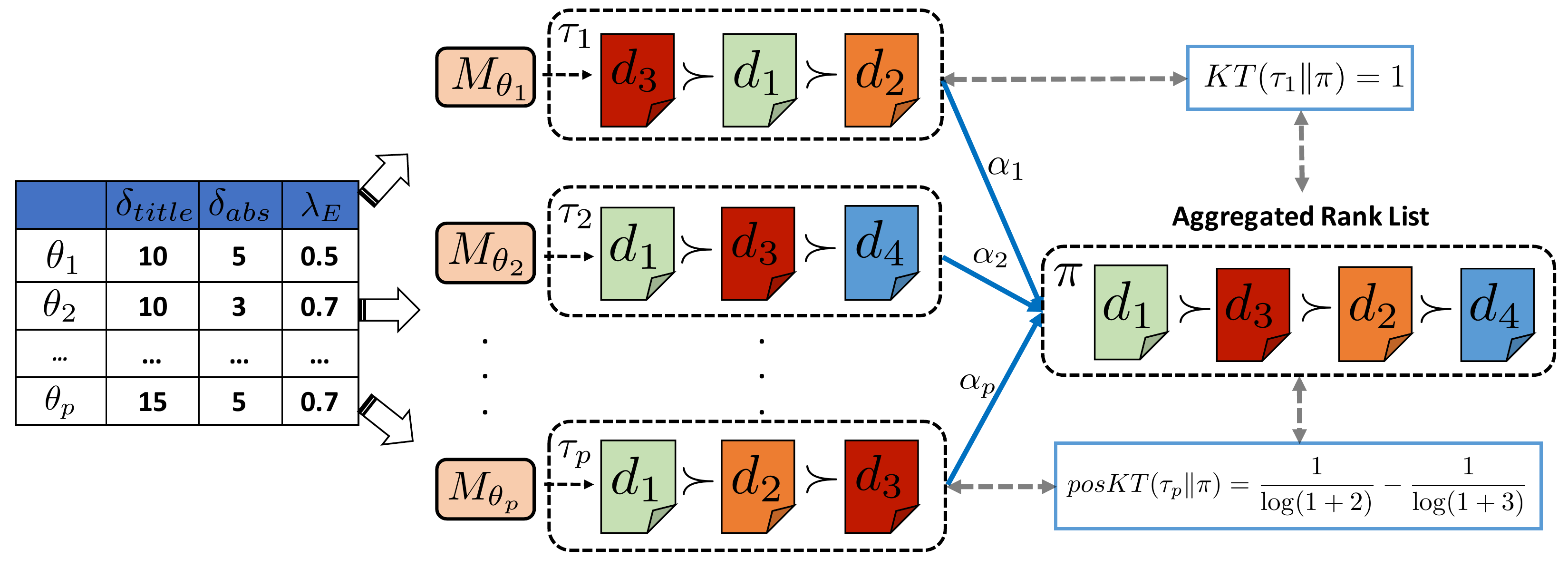}
  \caption{\small An illustrative example showing the process of weighted rank aggregation and the calculation of two different ranking distances (\ie, $KT$ and $posKT$).}\label{fig:ModelSelectionExample}
  \vspace{-0.5em}
\end{figure}
Our framework measures the goodness of each parameter settings $\theta_{i} \in \Theta$ based on its induced ranking model $M_{\theta_{i}}$. 
The key challenge here is how we can evaluate the ranking performance of each $M_{\theta_{i}}$ over a query $q$ which has no labeled documents. 
To address this challenge, we first leverage a weighted rank aggregation technique to obtain an aggregated rank list and then evaluate the quality of each $M_{\theta_{i}}$ based on the agreement between its generated rank list and the aggregated rank list. 
The key intuition here is that high-quality ranking models will rank documents based on a similar distribution while low-quality ranking models will rank documents in a uniformly random fashion. 
Therefore, the agreement between each rank list with the aggregated rank list serves as a good signal of its quality.

At a high level, our model selection method is an iterative algorithm which repeatedly aggregates multiple rankings (with their corresponding weights) and uses the aggregated rank list to estimate the quality of each of them.
Given a query $q$, we first construct $p$ ranking models $M_{\theta_{i}},  i \in [1,\dots,p]$, one for each parameter settings $\theta_{i} \in \Theta$ and obtain its returned top-$k$ rank list $\tau_{i}$ over a document set $D_{i}$ (\ie, $|D_{i}| = k$). 
Then, we construct a unified document pool $D = \bigcup_{i=1}^{p} D_{i}$.
After that, we use $\alpha_{i}$ to denote the confidence score of each ranking model $M_{\theta_{i}}$, and initialize all of them with equal value $\frac{1}{p}$.
During each iteration, we first aggregate $\{\tau_{1}, \dots, \tau_{p}\}$, weighted by $\{\alpha_{1}, \dots, \alpha_{p}  \}$, and obtain the aggregated rank list $\pi$. 
Then, we adjust the confidence score of each ranking model $M_{\theta_{i}}$ (\ie, $\alpha_{i}$) based on the distance of two rankings: $\tau_{i}$ and $\pi$. 
Here, we use $\pi$ to denote the aggregated rank list because it is a complete ranking over the document pool $D$. 

%Given a query $q$, for each , we can  and obtain a rank list over document set $D_{i}$. In most cases, we will retrieve only the top-$k$ relevant results and thus we have $|D_{i}| = k$. After that, we construct a document pool $D = \bigcup_{i}^{p} D_{i}$, with size $|D| > k$ in general. 
%Therefore, the rank list for each ranking model $M_{\theta_{i}}$ is an incomplete ranking and we use $\tau_{i}$ to denote it. 
%Furthermore, we use $\alpha_{i}$ to denote the confidence score of each ranking model $M_{\theta_{i}}$, and initialize all of them with equal value $\frac{1}{p}$.
%Then, we apply our iterative model selection framework to update each $\alpha_{i}$.
%During each iteration, we first aggregate $\{\tau_{1}, \dots, \tau_{p}\}$, weighted by $\{\alpha_{1}, \dots, \alpha_{p}  \}$, and obtain the aggregated rank list $\pi$. 
%Here, we use $\pi$ because it is a complete ranking over the document pool $D$. 
%Then, we adjust the confidence score $\alpha_{i}$ for ranking model $M_{\theta_{i}}$ based on the distance of two rankings: $\tau_{i}$ and $\pi$. 

\vspace{0.1cm}
\noindent \textbf{Weighted Rank Aggregation.}
We aggregate multiple rank lists using a variant of Borda counting method \cite{coppersmith2006ordering} which considers the relative weight of each rank list. 
We calculate the score of each document based on its position in each rank list as follows:
\small
\begin{equation}
score(d_{j}) = \sum_{i=1}^{p} \alpha_{i} \left(|\tau_{i}|+1 - \tau_{i}(d_j) \right) \mathbf{1}\{ j \in I(\tau_{i})\},
\end{equation}
\normalsize
where $|\tau_{i}|$ denotes the length of a rank list $\tau_{i}$, and $\mathbf{1}\{x\}$ is an indicator function. 
When document $d_{j}$ appears in the rank list $\tau_{i}$, $\mathbf{1}\{ j \in I(\tau_{i})\}$ equals to 1, otherwise, it equals to 0. 
The above equation will reward a document ranked at higher position (\ie, small $\tau_{i}(d_j)$) in a high-quality rank list (\ie, large $\alpha_{i}$) a larger score.
Finally, we obtain the aggregated rank of these documents based on their corresponding scores. 
A concrete example in shown in Figure \ref{fig:ModelSelectionExample}.

\vspace{0.1cm}
\noindent \textbf{Confidence Score Adjustment.}
After we obtain the aggregated rank list, we will need to adjust the confidence score $\alpha_{i}$ of each ranking model $M_{\theta_{i}}$ based on the distance between $\tau_{i}$ and aggregated rank list $\pi$. 
In order to compare the distance between an incomplete rank list $\tau_{i}$ with a complete rank list $\pi$, we extend the classical Kendall Tau distance \cite{kendall1955rank} and define it as follows:
\small
\begin{equation}\label{eq:KT}
KT(\tau_{i} || \pi) = \sum_{\substack{\tau_{i}(a) < \tau_{i}(b)\\ a,b \in I(\tau_{i}) }} \mathbf{1} \{\pi(a) > \pi(b)\}. 
\end{equation}
\normalsize
The above distance counts the number of pairwise disagreements between $\tau_{i}$ and $\pi$. 
One limitation of this distance is that it does not differentiate the importance of different ranking positions. 
Usually, switching two documents in the top part of a rank list should be penalized more, compared with switching another two documents in the bottom part of a rank list. 
To model such intuition, we propose a \emph{position-aware} Kendall Tau distance and define it as follows:
\scriptsize
\begin{align}\label{eq:dKT}
\begin{aligned}
posKT(\tau_{i} || \pi) & =   \sum_{\substack{\tau_{i}(a) < \tau_{i}(b)\\ a,b \in I(\tau_{i}) }} \left(\frac{1}{\log_{2}(1 + \pi(b))} - \frac{1}{\log_{2}(1 + \pi(a))} \right)  \mathbf{1}\{\pi(a) > \pi(b)\}.
\end{aligned}
\end{align}
\normalsize
With the distance between two rankings defined, we can adjust the confidence score as follows:
\small
\begin{equation}\label{eq:adjust_alpha}
\alpha_{i} = \frac{\exp(-dist(\tau_{i} || \pi))}{\sum_{i'} \exp(-dist(\tau_{i'} || \pi)) },
\end{equation}
\normalsize
where $dist(\tau_{i} || \pi)$ can be either $KT(\tau_{i} || \pi)$ or $posKT(\tau_{i} || \pi)$ and we will study how different this choice can influence the model selection results in Section \ref{subsec:model_selection_results}. The key idea of the above equation is to promote the ranking model which returns a ranked list better aligned with the aggregated rank list.

\vspace{0.1cm}
\noindent \textbf{Putting all together.}
Algorithm \ref{alg:unsupervised_model_selection} summarizes our unsupervised model selection process. 
Given a query $q \in Q$, we can iteratively apply weighted rank aggregation and confidence score adjustment until the algorithm converges. 
Then, we collect the converged $\{\hat{\alpha_{1}}, \dots, \hat{\alpha_{p}}\}$. 
Specifically, $\hat{\alpha_{i}}$ is the confidence score of ranking model $M_{\theta_{i}}$ on query $q$. 
With a slight abuse of notation, we use $score(M_{\theta_{i}})$ to denote its accumulated confidence score.
Given a set of queries $Q$, we run the former procedure for each query and sum over all converged $\hat{\alpha_{i}}$. 
Finally, we return the ranking model $M_{\theta^{*}}$ which has the largest accumulated confidence score.

%!TEX root = main.tex
% UTF-8 encoding
\section{Experiments}\label{sec:exp}

In this section, we evaluate our proposed \SetRank~framework as well as unsupervised model selection algorithm on two datasets from two scientific domains. 

%% Dataset description
\vspace{-0.2ex}
\subsection{Datasets}\label{subsec:data}

We use two benchmark datasets\footnote{\scriptsize Both benchmark datasets are publicly available at: \url{https://github.com/mickeystroller/SetRank}.} for the experiments: \textsf{Semantic Scholar}~\cite{Xiong2017ExplicitSR} in \underline{C}omputer \underline{S}cience (\textsf{S2-CS}) and \textsf{TREC} 2004\&2005 Genomics Track in \underline{Bio}medical science (\textsf{TREC-BIO}). 

\noindent \textbf{\textsf{S2-CS}} contains 100 queries sampled from \textsf{Semantic Scholar}'s query log, in which 40 queries are entity-set queries and the maximum number of entities in a query is 5. 
Candidate documents are generated by pooling from variations of \textsf{Semantic Scholar}'s online production system and all of them are manually labeled on a 5-level scale. 
Entities in both queries and documents are linked to Freebase using \textsf{CMNS}~\cite{hasibi2015entity}. 
As the original dataset does not contain the entity type information, we enhance it by retrieving each entity's most notable type in the latest Freebase dump\footnote{\scriptsize \url{https://developers.google.com/freebase/}} based on its Freebase ID. These types are organized by Freebase type hierarchy. 

\noindent \textbf{\textsf{TREC-BIO}} includes 100 queries designed by biologists and the candidate document pool is constructed based on the top results of all submissions at that time. 
All candidate documents are labeled on a 3-level scale. 
In these 100 queries, 86 of them are entity-set queries and the maximum number of entities in a query is 11. 
The original dataset contains no entity information and therefore we apply PubTator~\cite{Wei2013PubTatorAW}, the state-of-the-art biomedical entity linking tool, to obtain 5 types of entities (\ie, \textit{Gene}, \textit{Disease}, \textit{Chemical}, \textit{Mutation}, and \textit{Species}) in both queries and documents. 
We build a simple type hierarchy with root node named \textsf{`Thing'} and each first-level node corresponds to one of the above 5 types.

%% Entity Linking performance
\vspace{-0.2ex}
\subsection{Entity Linking Performance}\label{subsec:entity-linking-performance}

We evaluate the query entity linking using precision and recall at the query level. 
Specifically, an entity annotation is considered correct if it appears in the gold labeled data (\ie, the strict evaluation in \cite{Carmel2014ERD14ER}). 
The original \textsf{S2-CS} dataset provides such gold labeled data. 
For \textsf{TREC-BIO} dataset, we asked two Master-level students with biomedical science background to label all the linked entities as well as the entities that they could identify in the queries. 
We also report the entity linking performance on the general domain queries (\textsf{ClueWeb09} and \textsf{ClueWeb12}) for references \cite{Xiong2016BagofEntitiesRF}. 
As we can see in Table \ref{tbl:entity-linking-performance}, the overall linking performance of academic queries is better than that of general domain queries, probably because academic queries have less ambiguity. 
Also, recall of entity linking in \textsf{TREC-BIO} dataset is very high. 
A possible reason is that the biomedical entities have very distinctive tokens (\eg, \mquote{narcolepsy} is a specific disease related to sleep and is seldom used in other contexts) and thus it is relatively easier to recognize them. 

%% Ranking Performance
\subsection{Ranking Performance}

%%% 1. Experimental Setup 
\subsubsection{Experimental Setup}
\hfill

\noindent\textbf{Evaluation metrics.} 
Since documents in both datasets have multi-level graded relevance, we use NDCG@\{5,10,15,20\} as our main evaluation metrics. 
All evaluation is performed using standard \textsf{pytrec\_eval} tool \cite{VanGysel2018pytreceval}.
Statistical significances are tested using two-tailed $t$-test with p-value $\leq 0.05$. 

\noindent\textbf{Baselines.} 
We compare \SetRank~with 4 baseline ranking models: Vector Space Model (BM25~\cite{Robertson2009ThePR}), Query Likelihood Model with Dirichlet Prior smoothing (LM-DIR) or with Jelinek Mercer smoothing (LM-JM)~\cite{Zhai2001ASO}, and the Information-Based model (IB)~\cite{Clinchant2010InformationbasedMF}. 
All models are applied to the paper's title and abstract fields. 
Here, we do not compete with \textsf{Semantic Scholar}'s production system and ESR model~\cite{Xiong2017ExplicitSR} because they are supervised models trained over user's click information which is not available in our setting. 

The parameters of all models, including the field weights, are set using 5-fold cross validation over the queries in each benchmark dataset using the same paradigm in \cite{Raviv2016DocumentRU} as follows.
For each hold-out fold, the other four folds are served as a validation set.
A grid search is applied to choose the optimal parameter settings that maximize NDCG@20 on the validation set. 
Specifically, the title and abstract field weights are selected from \{1,5,10,15,20,50\}; 
the Dirichlet smoothing parameter $\mu$ and Jelinek Mercer smoothing parameter $\lambda$ are chosen from \{500, 1000, 1500, 2000, 2500, 3000\} and \{0.1, 0.2, \dots, 0.9\}, respectively;
the relative weight of entity token $\lambda_{E}$ used in \SetRank~is selected from \{0, 0.1, \dots, 1\}. 
The best performing parameter settings are then saved for the hold-out evaluation.

%% Table 2: Entity linking performance
        \begin{table}[!t]
            \caption{\small Entity linking performance on scientific domain queries (\textsf{S2-CS}, \textsf{TREC-BIO}) and general domain queries (\textsf{ClueWeb09}, \textsf{ClueWeb12}).}
            \label{tbl:entity-linking-performance}
            \vspace{-0.3cm}
	\scalebox{0.72}{
            \begin{tabular}{|c||c|c||c|c|}
                \hline
                 & \textbf{\textsf{S2-CS}}  & \textbf{\textsf{TREC-BIO}} & \textbf{\textsf{ClueWeb09}} & \textbf{\textsf{ClueWeb12}}  \\
                \hline
                \hline
                \textbf{Precision} & 0.680 & 0.678 & 0.577 & 0.485  \\
                \hline
                \textbf{Recall} & 0.680 & 0.727 & 0.596 & 0.575  \\
                \hline
            \end{tabular}
}
            \vspace{-0.5cm}
       \end{table}
       
%%% 2. Effectiveness of Leveraging Entity Information
\subsubsection{Effectiveness of Leveraging Entity Information}
\hfill

As mentioned before, the entity linking process is not perfect and it generates some noisy entity annotations. 
Therefore, we first study how different ranking models, including our proposed \SetRank, can leverage such noisy entity information to improve the ranking performance. 
We evaluate three variations of each model -- one using only word information, one using only entity information, and one using both pieces of information. 

Results are shown in Table~\ref{tbl:entity4ranking}. 
We notice that the usefulness of entity information is \emph{inconclusive} for baseline models. 
On \textsf{S2-CS} dataset, adding entity information can improve the ranking performance, while on \textsf{TREC-BIO} dataset, it will drag down the performance of all baseline methods. 
This resonates with previous findings in \cite{karimi2012quantifying} that simply adding entities into queries and posting them to existing ranking models does not work for biomedical literature retrieval. 
Compared with baseline methods, \SetRank~successfully combines the word and entity information and effectively leverages such noisy entity information to improve the ranking performance. 
Furthermore, \SetRank~can better utilize each single information source, either word or entity, than other baseline models thanks to our proposed query graph representation. 
Overall, \SetRank~significantly outperforms all variations of baseline models. 

%% Table 3: Word-Entity-Both-fives-methods
\begin{table*}[!thbp]
\caption{\small Effectiveness of leveraging (noisy) entity information for ranking. Each method contains three variations and the best variation is labeled bold. 
The superscript ``$*$" means the model significantly outperforms the best variation of all 4 baseline methods (with p-value $\leq$ 0.05). 
}
\label{tbl:entity4ranking}
\centering
\newcommand{\tabincell}[2]{\begin{tabular}{@{}#1@{}}#2\end{tabular}}
            \vspace{-0.3cm}
\scalebox{0.72}{
\begin{tabular}{|c|c||  c|c|c| c|c|c| c|c|c| c|c|c| c|c|c| }
\cline{1-17}
 &  & \multicolumn{3}{c|}{ BM25}   &  \multicolumn{3}{c|}{ LM-DIR} & \multicolumn{3}{c|}{ LM-JM} & \multicolumn{3}{c|}{ IB} & \multicolumn{3}{c|}{ \SetRankFull } \\
 \cline{1-17}
Dataset & Method &  \tabincell{c}{\emph{Word}} &  \tabincell{c}{\emph{Entity}} &  \tabincell{c}{\emph{Both}} &  \tabincell{c}{\emph{Word}} &  \tabincell{c}{\emph{Entity}} &  \tabincell{c}{\emph{Both}} &  \tabincell{c}{\emph{Word}} &  \tabincell{c}{\emph{Entity}} &  \tabincell{c}{\emph{Both}} &  \tabincell{c}{\emph{Word}} &  \tabincell{c}{\emph{Entity}} &  \tabincell{c}{\emph{Both}} &  \tabincell{c}{\emph{Word}} &  \tabincell{c}{\emph{Entity}} &  \tabincell{c}{\emph{Both}} \\
\hline
\hline
\multirow{4}{*}{\makecell{\textsf{S2-CS}}} 
&  NDCG@5 &  0.3476 & 0.3319 & \textbf{0.3675} & 0.3447 & 0.3460 & \textbf{0.3563} & \textbf{0.3626} & 0.3394 & 0.3625 & \textbf{0.3759} & 0.3420 & 0.3729 & 0.3890 & 0.3761 & \textbf{0.4207}$^{*}$ \\
\cline{2-17}
 &  NDCG@10 & 0.3785 & 0.3520 & \textbf{0.4039} & 0.3623 & 0.3579 & \textbf{0.3901} & 0.3774 & 0.3519 & \textbf{0.3962} & 0.3903 & 0.3557 & \textbf{0.4009} & 0.4168 & 0.3885 & \textbf{0.4431}$^{*}$ \\
 \cline{2-17}
 &  NDCG@15 & 0.4001 & 0.3616 & \textbf{0.4160} & 0.3781 & 0.3673 & \textbf{0.4077} & 0.4051 & 0.3666 & \textbf{0.4174} & 0.4113 & 0.3699 & \textbf{0.4272} & 0.4411 & 0.4054 & \textbf{0.4762}$^{*}$ \\
\cline{2-17}
 &  NDCG@20 & 0.4126 & 0.3752 & \textbf{0.4333} & 0.4012 & 0.3816 & \textbf{0.4205} & 0.4182 & 0.3804 & \textbf{0.4362} & 0.4295 & 0.3855 & \textbf{0.4421} & 0.4674 & 0.4229 & \textbf{0.4950}$^{*}$ \\
\cline{2-17}
\cline{1-1}
\hline
\hline
\multirow{4}{*}{\textsf{\makecell{TREC-BIO}}} 
&  NDCG@5 & \textbf{0.3189} & 0.1542 & 0.2613 & \textbf{0.3053} & 0.1755 & 0.2669 & \textbf{0.2957} & 0.1656 & 0.2826 & \textbf{0.3045} & 0.1842 & 0.2770  & 0.3417 & 0.2111 & \textbf{0.3744}$^{*}$ \\
\cline{2-17}
 &  NDCG@10 & \textbf{0.2968} & 0.1488 & 0.2472 & \textbf{0.2958} & 0.1601 & 0.2571 & \textbf{0.2742} & 0.1588 & 0.2572 & \textbf{0.2918} & 0.1715 & 0.2633 & 0.3165 & 0.1976 & \textbf{0.3522}$^{*}$ \\
\cline{2-17}
 &  NDCG@15 & \textbf{0.2833} & 0.1424 & 0.2395 & \textbf{0.2852} & 0.1579 & 0.2591 & \textbf{0.2642} & 0.1575 & 0.2437 & \textbf{0.2835} & 0.1664 & 0.2541 & 0.3017 & 0.1931 & \textbf{0.3363}$^{*}$ \\
\cline{2-17}
 &  NDCG@20 & \textbf{0.2739} & 0.1419 & 0.2337 & \textbf{0.2781} & 0.1558 & 0.2547 & \textbf{0.2560} & 0.1534 & 0.2362 & \textbf{0.2722} & 0.1628 & 0.2406 & 0.2900 & 0.1885 & \textbf{0.3246}$^{*}$ \\
\cline{2-17}
\cline{1-1}
\end{tabular}
}
\vspace{-0.3cm}
\end{table*}

%%% 3. Performance on Entity-Set Query
\subsubsection{Ranking Performance on Entity-Set Queries}
\hfill

We further study each model's ranking performance on entity-set queries.
There are 40 and 86 entity-set queries in \textsf{S2-CS} and \textsf{TREC-BIO}, respectively. 
We denote these subsets of entity-set queries as \textsf{S2-CS-ESQ} and \textsf{TREC-BIO-ESQ}. 
As shown in Table~\ref{tbl:esq-ranking}, \SetRank~significantly outperforms the best variation of all baseline methods on \textsf{S2-CS-ESQ} and \textsf{TREC-BIO-ESQ} by at least 25\% and 14\% respectively in terms of NDCG@5. 
Also, we can see the advantages of \SetRank~over the baselines on entity-set queries are larger than those on general queries, 
This further demonstrates \SetRank's effectiveness of modeling entity set information. 

%%% Table 4: five methods over ESQ in two dataset
\begin{table}[t]
\caption{\small Ranking performance on entity-set queries. The best variation of each baseline method is selected. 
The superscript ``$*$" means the model significantly outperforms all 4 baseline methods (with p-value $\leq$ 0.05). 
}
\label{tbl:esq-ranking}
\centering
            \vspace{-0.3cm}
\scalebox{0.72}{
\begin{tabular}{|c|c|| c|c|c|c|c|}
\cline{1-7}
Dataset & Metric &  BM25 & LM-DIR & LM-JM & IB & \SetRankFull \\
\hline\hline
\multirow{4}{*}{\textsf{\makecell{S2-CS\\-ESQ}}} & 
NDCG@5 & 0.3994 & 0.3522 & 0.3812 & 0.3956 & \textbf{0.4983}$^{*}$ \\
\cline{2-7}
 & NDCG@10 & 0.4364 & 0.3973 & 0.4241 & 0.4209 & \textbf{0.5130}$^{*}$ \\
\cline{2-7}
 & NDCG@15 & 0.4454 & 0.4160 & 0.4431 & 0.4496 &  \textbf{0.5450}$^{*}$ \\
\cline{2-7}
 & NDCG@20 & 0.4609 & 0.4264 & 0.4618 & 0.4664 &  \textbf{0.5629}$^{*}$ \\
\cline{2-7}
\cline{1-1}
\hline
\hline
\multirow{4}{*}{\textsf{\makecell{TREC-BIO\\-ESQ}}} & 
NDCG@5  & 0.3185 & 0.2934 & 0.2940 & 0.3011 & \textbf{0.3639}$^{*}$ \\
\cline{2-7}
 & NDCG@10 & 0.2968 & 0.2834 & 0.2746 & 0.2896 & \textbf{0.3406}$^{*}$ \\
\cline{2-7}
 & NDCG@15 & 0.2812 & 0.2711 & 0.2636 & 0.2832 & \textbf{0.3251}$^{*}$ \\
\cline{2-7}
 & NDCG@20 & 0.2718 & 0.2644 & 0.2553 & 0.2708 & \textbf{0.3132}$^{*}$ \\
\cline{2-7}
\cline{1-1}
\hline
\end{tabular}
}
            \vspace{-0.3cm}
\end{table}

%%% 4. Effectiveness of Modeling Entity Relation and Entity Type
\subsubsection{Effectiveness of Modeling Entity Relation and Entity Type}
\hfill

To study how the inter-entity relation and entity type information can contribute to document ranking, we compare \SetRank~with two of its variants, \SetRankNoType~and \SetRankNoTypeNoSet.
The first variant models entity relation among the set but ignores the entity type information, and the second variant simply neglects both entity relation and type. 

Results are shown in Table~\ref{tbl:setrank-variation-performance}. 
First, we compare \SetRankNoType~with \SetRankNoTypeNoSet~and find that modeling the entity relation in entity sets can significantly improve the ranking results. 
Such improvement is especially obvious on the entity-set query sets \textsf{S2-CS-ESQ} and \textsf{TREC-BIO-ESQ}. 
Also, by comparing \SetRank~with \SetRankNoType, we can see adding entity type information can further improve ranking performance. 
In addition, we present a concrete case study for one entity-set query in Table \ref{tbl:case1}. 
The top-2 papers returned by \SetRankNoTypeNoSet~are focusing on \emph{video game} without discussing its relation with \emph{reinforcement learning}.
In comparison, \SetRank~considers the entity relations and returns the paper mentioning both entities.

%%% Table 5: different variations of SetRank over 4 datasets 
\begin{table}[!htbp]
\caption{\small Ranking performance of different variations of \SetRank. 
Best results are marked bold. 
The superscript ``$*$" means the model significantly outperforms \SetRankNoTypeNoSet~(with p-value $\leq$ 0.05). 
}\label{tbl:setrank-variation-performance}
\centering
\vspace{-0.3cm}
\scalebox{0.75}{
\begin{tabular}{|c|c|| c|c|c|}
\cline{1-5}

Dataset & Metric & \SetRankNoTypeNoSet & \SetRankNoType & \SetRankFull \\
\hline\hline

\multirow{4}{*}{\makecell{\textsf{S2-CS}}} & 
NDCG@5 & 0.3847 & 0.4157$^{*}$ & \textbf{0.4207}$^{*}$ \\
\cline{2-5}
 & NDCG@10 & 0.4095 & 0.4423$^{*}$ & \textbf{0.4431}$^{*}$ \\
\cline{2-5}
 & NDCG@15 & 0.4256 & 0.4655$^{*}$ & \textbf{0.4762}$^{*}$  \\
\cline{2-5}
 & NDCG@20 & 0.4443 & 0.4813$^{*}$ & \textbf{0.4950}$^{*}$  \\
\cline{2-5}
\cline{1-1}
\hline
\hline
\multirow{4}{*}{\makecell{\textsf{TREC-BIO}}} & 
NDCG@5 & 0.3414 & 0.3705 & \textbf{0.3744} \\
\cline{2-5}
 & NDCG@10 & 0.3257 & 0.3500 & \textbf{0.3522}$^{*}$ \\
\cline{2-5}
 & NDCG@15 & 0.3140 & 0.3335 & \textbf{0.3363}$^{*}$ \\
\cline{2-5}
 & NDCG@20 & 0.3058 & 0.3217 & \textbf{0.3246} \\
\cline{2-5}
\cline{1-1}
\hline
\hline
\multirow{4}{*}{\textsf{\makecell{S2-CS\\-ESQ}}} & 
NDCG@5 & 0.4059 & 0.4800$^{*}$ & \textbf{0.4983}$^{*}$ \\
\cline{2-5}
 & NDCG@10 & 0.4311 & 0.5004$^{*}$ & \textbf{0.5130}$^{*}$ \\
\cline{2-5}
 & NDCG@15 & 0.4469 & 0.5266$^{*}$ & \textbf{0.5450}$^{*}$ \\
\cline{2-5}
 & NDCG@20 & 0.4683 & 0.5378$^{*}$ & \textbf{0.5629}$^{*}$ \\
\cline{2-5}
\cline{1-1}
\hline
\hline
\multirow{4}{*}{\textsf{\makecell{TREC-BIO\\-ESQ}}} & 
NDCG@5 & 0.3257 & 0.3594 & \textbf{0.3639}$^{*}$ \\
\cline{2-5}
 & NDCG@10 & 0.3100 & 0.3380$^{*}$ & \textbf{0.3406}$^{*}$ \\
\cline{2-5}
 & NDCG@15 & 0.2994 & 0.3219$^{*}$ & \textbf{0.3251}$^{*}$ \\
\cline{2-5}
 & NDCG@20 & 0.2903 & 0.3100$^{*}$  & \textbf{0.3132}$^{*}$  \\
\cline{2-5}
\cline{1-1}
\hline
\end{tabular}
}
\vspace{-0.3cm}
\end{table}

%% Table: case studies on S2-CS dataset. 
\begin{table*}[!htbp]
\centering
\caption{\small A case study comparing \SetRank~with \SetRankNoTypeNoSet~on one entity-set query in \textsf{S2-CS}. Note: \emph{Atari} is a video game platform.}
\label{tbl:case1}
\vspace{-0.3cm}
\scalebox{0.65}{
\begin{tabular}{|c|| c|c|}
\hline
Query  &  \multicolumn{2}{c|}{ \textbf{reinforcement learning for video game}}  \\ \hline
Method & \SetRankNoTypeNoSet                                                                                      & \SetRank                                                                  \\ 
\hline
1      & The effects of \textcolor{red}{video game} playing on attention, memory, and executive control                           & A \textcolor{red}{video game} description language for model-based or \textcolor{orange}{interactive learning}   \\ \hline
2      & Can training in a real-time strategy \textcolor{red}{video game} attenuate cognitive decline in older adults?        & Playing \textcolor{red}{Atari} with Deep \textcolor{orange}{Reinforcement Learning}  \\ 
\hline
3      & A \textcolor{red}{video game} description language for model-based or \textcolor{orange}{interactive learning}   &     Real-time \textcolor{orange}{neuroevolution} in the NERO \textcolor{red}{video game}  \\ 
\hline
\end{tabular}
}
\vspace{-0.2cm}
\end{table*}

%% 5. Parameter Sensitivity Analysis
\subsubsection{Analysis of Entity Token Weight $\lambda_{E}$}
\hfill 

We introduce the entity token weight $\lambda_{E}$ in Eq. (\ref{eq:node_lambda}) to combine the entity-based and word-based relevance scores. 
In all previous experiments, we choose its value using cross validation. 
Here, we study how this parameter will influence the ranking performance by constructing multiple \SetRank~models with different $\lambda_{E}$ and directly report their performance on all 100 queries. 
\begin{figure}[!h]
\vspace{-0.2cm}
  \begin{center}
  \includegraphics[width=0.20\textwidth]{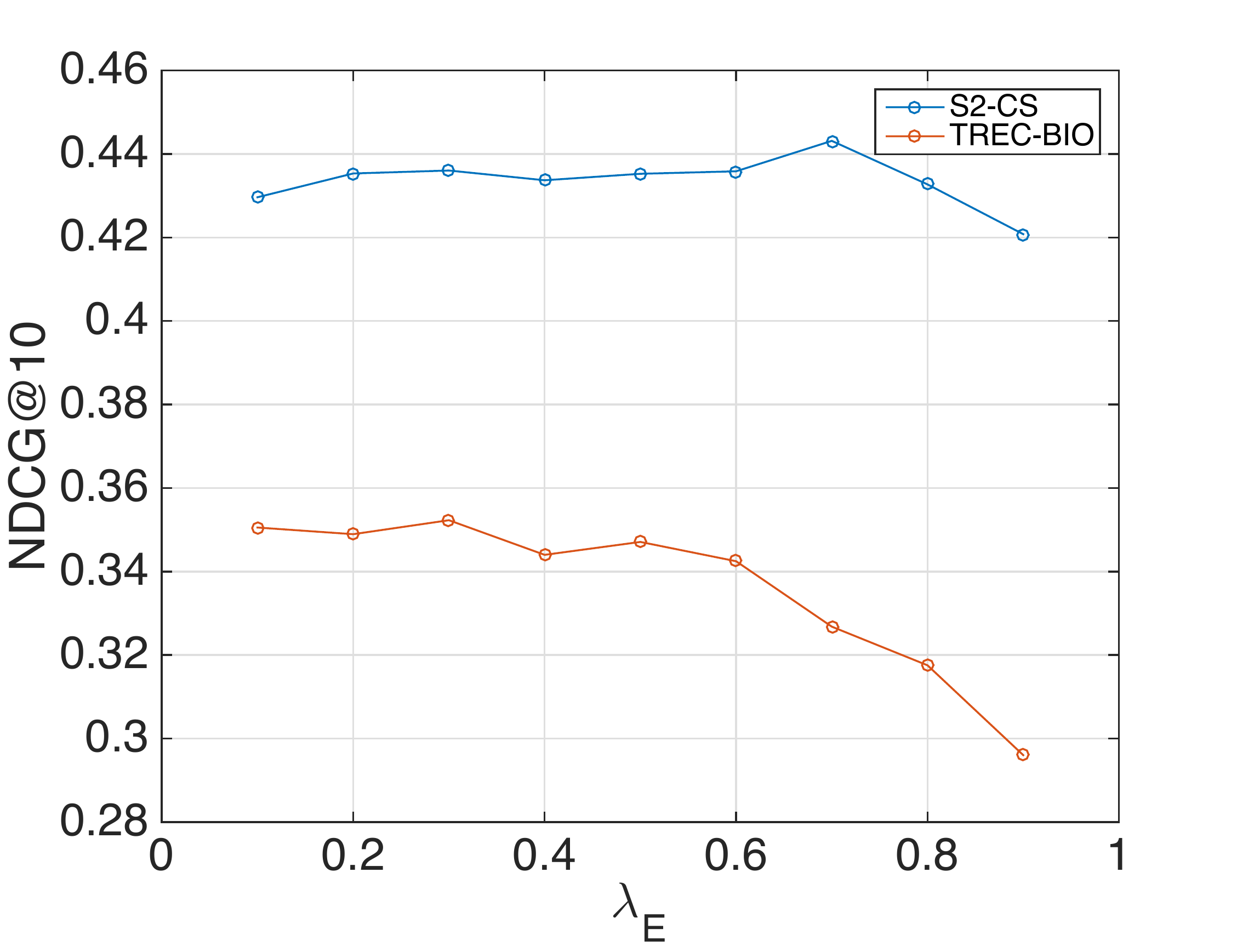}
  \end{center}
  \vspace{-0.4cm}
  \caption{\small Sensitivity of $\lambda_{E}$ in \textsf{S2-CS} and \textsf{TREC-BIO} datasets.}
  \label{subfig:lambda_sensitivity}
  \vspace{-0.3cm}
\end{figure}

As shown in Figure \ref{subfig:lambda_sensitivity}, for \textsf{S2-CS} dataset, \SetRank's ranking performance first increases as $\lambda_{E}$ increases until it reaches 0.7 and then starts to decrease when we further increase $\lambda_{E}$. 
However, for \textsf{TREC-BIO} dataset, the optimal value of $\lambda_{E}$ is around 0.3, and if we increases $\lambda_{E}$ over 0.6, the ranking performance will drop quickly. 

%% Table: Effectiveness of ranking model selection
\begin{table*}[!htbp]
\caption{\small Effectiveness of ranking model selection. 
\SetRank-\textsf{$VS$}: parameters are tuned using 5-fold cross validation. 
\AutoSetRank-\textsf{($KT/posKT$)}: parameters are obtained based on our unsupervised model selection algorithm, which uses either \underline{$KT$} or \underline{$posKT$} as ranking distance.
Mean ($\pm$ Std): the averaged performance of all ranking models with standard derivation shown. 
}
\label{tbl:model-selection}
\centering
\vspace{-0.3cm}
\scalebox{0.75}{
\begin{tabular}{|c|c|| @{\hskip1pt}c@{\hskip1pt}| @{\hskip1pt}c@{\hskip1pt}| c| @{\hskip1pt}c@{\hskip1pt}| @{\hskip1pt}c@{\hskip1pt}| c|c|c|c|}
\cline{1-11}
Dataset & Method & $\delta_{title}$ & $\delta_{abs}$ & $\lambda_{E}$ & $\mu_{title}$ & $\mu_{abs}$ & NDCG@5 & NDCG@10 & NDCG@15 & NDCG@20 \\
\hline\hline
\multirow{4}{*}{\textsf{\makecell{S2-CS}}} & 
\SetRank-\textsf{$VS$} & 20 & 5 & 0.7 & 1000 & 1000 & 0.4207 & 0.4431 & 0.4762 & 0.4950 \\
\cline{2-11}
 & \AutoSetRank-\textsf{$KT$} & 20 & 7 & 0.7 & 1500 & 2000  &  0.4174 &  0.4427 & 0.4730  &  0.4929 \\
\cline{2-11}
 & \AutoSetRank-\textsf{$posKT$} & 20 & 5 & 0.7 & 1500 & 1500  &  0.4173 &  0.4436 &  0.4731 & 0.4923  \\
\cline{2-11}
 & Mean ($\pm$ Std)  & -- & -- & -- & -- & -- &  0.3898 ($\pm$ 0.0112) & 0.4128 ($\pm$ 0.0106)  & 0.4411 ($\pm$ 0.0161 ) &  0.4543 ($\pm$ 0.0163) \\
\cline{2-11}
\cline{1-1}
\hline
\hline
\multirow{4}{*}{\textsf{\makecell{TREC-BIO}}} & 
\SetRank-\textsf{$VS$} & 20 & 5 & 0.2 & 1000 & 1000 & 0.3744 & 0.3522 & 0.3363 & 0.3246 \\
\cline{2-11}
 & \AutoSetRank-\textsf{$KT$}  & 20 & 5 & 0.2 & 1500 & 1000 &  0.3692 & 0.3472 & 0.3305 & 0.3173 \\
\cline{2-11}
 & \AutoSetRank-\textsf{$posKT$}  & 20 & 7 & 0.2 & 1000 & 1000 & 0.3748 & 0.3564 & 0.3367 &  0.3253 \\
\cline{2-11}
 & Mean ($\pm$ Std)  & -- & -- & -- & -- & -- & 0.3479 ($\pm$ 0.0103) & 0.3238 ($\pm$ 0.0079)  & 0.3199 ($\pm$ 0.0079) & 0.3036 ($\pm$ 0.0093) \\
\cline{2-11}
\cline{1-1}
\hline
\end{tabular}
}
\vspace{-0.3cm}
\end{table*}

%% Effectiveness of Model Selection
\subsection{Effectiveness of Model Selection}\label{subsec:model_selection_results}

\subsubsection{Experimental Setup}
\hfill 

In this experiment, we try to apply our unsupervised model selection algorithm to choose the best parameter settings of \SetRank~without using a validation set. 
We select entity token weight $\lambda_{E}$, title field weight $\delta_{title}$, abstract field weight $\delta_{abs}$, dirichlet smoothing factors for both fields $\mu_{title}$ \& $\mu_{abs}$ from \{0.2, 0.3, \dots, 0.8\}, \{5, 10, 15, 20\}, \{1, 3, 5, 10\}, and \{500, 1000, 1500, 2000\}, respectively. 
This generates totally $7 \times 4 \times 4 \times 4 \times 4 = 1,792$ possible parameter settings and for each of them we can construct a ranking model. 
We first apply our unsupervised model selection algorithm (with either $KT$ or $posKT$ as the ranking distance) and obtain the most confident parameter settings returned by it.
Then, we plug in these parameter settings into \SetRank~and denote it as \AutoSetRank. 
For reference, we also calculate the average performance of all 1,792 ranking models.

\subsubsection{Experimental Result and Analysis}
\hfill

Table \ref{tbl:model-selection} shows the results, including the \SetRank's performance when a labeled validation set is given. 
First, we notice that for \textsf{S2-CS} dataset, although the parameter settings tuned over validation set do perform better than the ones returned by our unsupervised model selection algorithm, the difference is not significant. 
For \textsf{TREC-BIO} dataset, it is surprising to find that \AutoSetRank-\textsf{$posKT$} can slightly outperforms \SetRank~tuned on validation set.
Furthermore, the performance of \AutoSetRank~function is higher than the average performances of all possible ranking models by 2 standard deviations, which demonstrates the effectiveness of our unsupervised model selection algorithm. 

%% Case Studies
\subsection{Use Case Study: Bio-Literature Search}

%% Table: Bio use case
\begin{table*}[!htbp]
\caption{\small A real-world use case comparing \SetRank~with PubMed. 
The input query contains a set of 10 genes and reflects user's information need of finding an association between this gene set and an unknown disease. 
Entity mentions in returned paper titles are highlighted in brown and the entity mentions of Alzheimer's disease, which are used to judge paper relevance, are marked in red. 
}
\vspace{-0.3cm}
\label{tbl:bio-case}
\centering
\scalebox{0.75}{
\begin{tabular}{|c| @{\hskip 0.5pt}c@{\hskip 0.5pt} ||c|}
\cline{1-3}
Query & \multicolumn{2}{c|}{ \textbf{APP APOE4 PSEN1 SORL1 PSEN2 ACE CLU BDNF IL1B MAPT}}  \\ \hline
Method & Rank & Paper Title \\ 
\hline\hline
\multirow{5}{*}{\makecell{PubMed}} & 1 & Apathy and \textcolor{brown}{APOE4} are associated with reduced \textcolor{brown}{BDNF} levels in \textcolor{red}{Alzheimer's disease} \\
\cline{2-3}
 & 2 & \textcolor{brown}{ApoE4} and A$\beta$ Oligomers Reduce \textcolor{brown}{BDNF} Expression via HDAC Nuclear Translocation \\
\cline{2-3}
 & 3 & Cognitive deficits and disruption of neurogenesis in a mouse model of \textcolor{brown}{apolipoprotein E4} domain interaction \\
\cline{2-3}
 & 4 & \textcolor{brown}{APOE-epsilon4} and aging of medial temporal lobe gray matter in healthy adults older than 50 years \\
\cline{2-3}
 & 5 & Influence of \textcolor{brown}{BDNF} Val66Met on the relationship between physical activity and brain volume \\
\cline{2-3}
\cline{1-1}
\hline
\hline
\multirow{5}{*}{\makecell{\SetRank}} & 1 & Investigating the role of rare coding variability in Mendelian dementia genes (\textcolor{brown}{APP}, \textcolor{brown}{PSEN1}, \textcolor{brown}{PSEN2}, GRN, \textcolor{brown}{MAPT}, and PRNP) in late-onset \textcolor{red}{Alzheimer's disease} \\
\cline{2-3}
 & 2 & Rare Genetic Variant in \textcolor{brown}{SORL1} May Increase Penetrance of \textcolor{red}{Alzheimer's Disease} in a Family with Several Generations of \textcolor{brown}{APOE- 4} Homozygosity \\
\cline{2-3}
 & 3 & \textcolor{brown}{APP}, \textcolor{brown}{PSEN1}, and \textcolor{brown}{PSEN2} mutations in early-onset \textcolor{red}{Alzheimer disease}: A genetic screening study of familial and sporadic cases \\
\cline{2-3}
 & 4 & Identification and description of three families with familial \textcolor{red}{Alzheimer disease} that segregate variants in the \textcolor{brown}{SORL1} gene \\
\cline{2-3}
 & 5 & The \textcolor{brown}{PSEN1}, p.E318G variant increases the risk of \textcolor{red}{Alzheimer's disease} in \textcolor{brown}{APOE-4} carriers \\
\cline{2-3}
\hline
\end{tabular}
}
\vspace{-0.3cm}
\end{table*}

In this section we demonstrate the effectiveness of \SetRank~in a biomedical use case. 
As preparation, we build a biomedical literature search engine based on over 27 million papers retrieved from PubMed. Entities in all papers are extracted and typed using PubTator. This search system is cold-started with our proposed \SetRank~model and we show how \SetRank~can help this search system to accommodate a given entity-set query and returns a high-quality rank list of papers relevant to the query. Comparison with PubMed, a widely used search engine for biomedical literature, will also be discussed.

\noindent \textbf{A biomedical case.} 
Consider the following case of a biomedical information need. 
Genomics studies often identify sets of genes as having important roles to play in the processes or conditions under investigation, and the investigators seek to understand better what biological insights such a list of genes might provide. 
Suppose such a study, having examined brain gene expression patterns in old mice, identifies ten genes as being of potential interest. 
The investigator forms a query with these 10 genes, submits it to a literature search engine, and examines the top ten returned papers to look for an association between this gene set and a disease. 
The query consists of symbols of the 10 genes: \mquote{APP, APOE, PSEN1, SORL1, PSEN2, ACE, CLU, BDNF, IL1B, MAPT}.

\noindent \textbf{Relevance criterion.} 
We choose the above ten genes for our illustration because these are actually top genes associated with Alzheimer's disease according to DisGeNET \cite{pinero2016disgenet}, and it is unlikely that there is another completely different (and unknown) commonality among them. 
Therefore, a retrieved paper is relevant if and only if it discusses at least one of the query genes in the context of Alzheimer's disease. Furthermore, among all relevant papers, we prefer those covering more unique genes.

\noindent \textbf{Result analysis.} 
The top-5 papers returned by PubMed\footnote{\scriptsize Querying PubMed with the exact same query returns 0 document. To get reasonable results, PubMed users have to insert an OR logic between every pairs of genes, and change the default ``sorting by most recent'' to ``sorting by best match''.} and our system are shown in Table \ref{tbl:bio-case}. 
We see that the \mquote{Alzheimer's disease} is explicitly mentioned in the title of all the five papers returned by our system, and the top two papers cover 6 unique genes among the total 10 genes. 
All five papers returned by \SetRank~are highly relevant, since they all focus on the association between a subset of our query genes and Alzheimer's disease.
In contrast, the top-5 papers retrieved by PubMed are dominated by two genes (\ie, APOE4 and BDNF) and contain none of the remaining eight. 
%Also, only one of these five papers mentions \mquote{Alzheimer's disease} explicitly in the title and two other papers (ranked 3rd and 5th) discuss some related concepts about \mquote{Alzheimer's disease}. 
%Moreover, the papers ranked 2nd and 4th are irrelevant. 
Only the 1st of the five papers is highly relevant. 
It focuses on the association between Alzheimer's disease (mentioned explicitly in the title) and our query gene set. 
Three other papers (ranked 2nd to 4th) are marginally relevant, in the sense that Alzheimer's disease is the context but not the focus of their studies. The paper ranked 5th is irrelevant.
Therefore, users will prefer \SetRank~since it returns papers covering a large-portion of an entity-set query and helps them to find the association between this entity set with Alzheimer's disease.

%!TEX root = main.tex
% UTF-8 encoding
\vspace{-0.1cm}
\section{Conclusions and Future Work}\label{sec:con}
In this paper, we study the problem of searching scientific literature using entity-set queries. 
A distinctive characteristic of entity-set queries is that they reflect user's interest in inter-entity relations.
To capture such information need, we propose \SetRank, an unsupervised ranking framework which explicitly models entity relations among the entity set. 
Second, we develop a novel unsupervised model selection algorithm based on weighted rank aggregation to select \SetRank's parameters without relying on a labeled validation set. 
Experimental results on two benchmark datasets corroborate the effectiveness of \SetRank~and the usefulness of our model selection algorithm. 
We further discuss the power of \SetRank with a real-world use case of biomedical literature search.

As a future direction, we would like to explore how we can go beyond pairwise entity relations and integrate higher-order entity relations into the current \SetRank~framework. 
Besides, it would be interesting to explore whether \SetRank~can effectively model domain expert's prior knowledge about the relative importance of entity relations.
Furthermore, the incorporation of user interaction and and extension of current \SetRank~framework to weakly-supervised settings are also interesting research problems. 

%!TEX root = main.tex
% UTF-8 encoding
\small
\section*{Acknowledgements}\label{sec:ack}
This research is sponsored in part by U.S. Army Research Lab. under Cooperative Agreement No. W911NF-09-2-0053 (NSCTA), DARPA under Agreement No. W911NF-17-C-0099, National Science Foundation IIS 16-18481, IIS 17-04532, and IIS-17-41317, DTRA HDTRA11810026, and grant 1U54GM114838 awarded by NIGMS through funds provided by the trans-NIH Big Data to Knowledge (BD2K) initiative (www.bd2k.nih.gov).
\normalsize

\bibliographystyle{ACM-Reference-Format}
\bibliography{cited}

\end{document}